\documentclass[aps. prl, reprint, superscriptaddress, floatfix, longbibliography, nobibnotes]{revtex4-2}

\usepackage[utf8]{inputenc}
\usepackage{amsmath}
\usepackage{amsfonts}
\usepackage{amssymb}
\usepackage{bbm}
\usepackage{siunitx}
\usepackage{physics}
\usepackage{stmaryrd}
\usepackage[colorlinks=true, citecolor=blue, urlcolor=red]{hyperref}
\usepackage{orcidlink}
\usepackage{svg}

\usepackage{cancel}
\usepackage[normalem]{ulem}

\usepackage{xr}

\newcommand{\mean}[1]{\langle {#1} \rangle}
\newcommand{\ii}{\mathbbm{i}}
\newcommand{\II}{ \mathbb I }
\newcommand{\E}{\mathrm{e}}
\newcommand{\D}{\mathrm{d}}
\newcommand{\p}[1]{\left({#1}\right)}
\newcommand{\pq}[1]{\left[{#1}\right]}

\definecolor{mlblue}{rgb}{0, 0.4470, 0.7410}
\definecolor{K6}{rgb}{0.3010, 0.7450, 0.9330}
\definecolor{K7}{rgb}{0.6350, 0.0780, 0.1840}
\definecolor{K8}{rgb}{0, 0.4470, 0.7410}
\definecolor{bp}{rgb}{1.0, 0.0, 0.5}

\newcommand{\alb}[1]{#1}

\newcommand{\e}{\mathrm{e}}

\newcommand{\cH}{\mathcal{H}}
\newcommand{\cU}{\mathcal{U}}
\newcommand{\cZ}{\mathcal{Z}}
\newcommand{\cF}{\mathcal{F}}

\newcommand{\trhou}{\tilde {\varrho}_1}
\newcommand{\trhoz}{\tilde {\varrho}_0}
\newcommand{\trhoD}{\tilde {\varrho}_\Delta}

\newcommand{\tPi}{\Pi^{\bot}}
\newcommand{\ttPi}{\tilde{\Pi}^{\bot}}
\newcommand{\Ut}{U_\tau}
\newcommand{\Utd}{U_\tau^\dagger}
\newcommand{\tUt}{\tilde{U}_\tau}
\newcommand{\tUtd}{\tilde{U}_\tau^\dagger}
\newcommand{\tp}{p^{\bot}}
\newcommand{\wtot}{w_{\mathrm{tot}}}
\newcommand{\wmeas}{w_{\mathrm{m}}}
\newcommand{\wex}{w_{\mathrm{ex}}}
\begin{document}
\title{Quantum fluctuation relations in first-detection processes }

\author{Alberto Imparato\,\orcidlink{0000-0002-7053-4732}}
\affiliation{Department of Physics, University of Trieste, Strada Costiera 11, 34151 Trieste, Italy}\email{alberto.imparato@units.it}
\affiliation{ Istituto Nazionale di Fisica Nucleare, Trieste Section, Via Valerio 2, 34127 Trieste, Italy}

\date{\today}

\begin{abstract}
	 We derive three quantum fluctuation relations for systems undergoing  repeated projective measurements. These fluctuation relations characterize the work that can be extracted from  a quantum device when a first-detection event triggers  a mechanical operation leading to positive work production.  The correction to the standard quantum Jarzynski equality depends logarithmically on the mean first-detection time for the time-reversed dynamics.  
	Application of  Jensen's inequality leads to fundamental limits on both the total work involved in the repeated measurements and  final mechanical operation, and  the extracted work alone. An analogous inequality is derived for the work performed exclusively during the sequence of repeated projective measurements.
The general case of a device connected to an external environment is also considered.
	
\end{abstract}

\maketitle
\section{ Introduction} 

Quantum thermodynamics provides a framework for investigating energy,
work, heat,  and information processing in systems operating
in the quantum regime \cite{Kosloff13,Anders16, Binder18, Deffner19}. One of its central objectives is to establish
fundamental bounds on thermodynamic quantities, particularly on the work
that can be extracted by quantum engines and feedback-controlled devices. In this context the possibility of designing and fabricating devices based on measurement feedback and capable of converting  information to work  has been
demonstrated experimentally in superconducting quantum systems
\cite{Masuyama2018,Naghiloo2018,Dassonneville2026}, while quantum
measurement backaction has been identified and exploited as an energetic
resource \cite{Elouard2018,Manikandan22}.

First-passage and stopping-time strategies offer a distinct approach to
feedback control. In these protocols, the action performed by the
controller depends not only on the measurement outcome, but also on the
first time at which a prescribed outcome is observed or a chosen threshold
is crossed. Stopping-time fluctuation relations have been derived for, e.g.,  gambling demons and  tested experimentally in  \cite{Manzano2021}. More recently, a classical
information engine in which the first crossing of a prescribed threshold
triggers work extraction has been experimentally realized
and a general fluctuation theorem for classical devices based on first-passage mechanism has been derived \cite{Archambault25}.

In the quantum domain, first-passage-time statistics have been formulated
for continuously monitored systems and measured experimentally using
repeated stroboscopic projective measurements
\cite{Kewming2024,Ryan2026}. These studies do not, however, implement a
thermodynamic operation triggered by the first-detection event. Conversely,
existing quantum Maxwell-demon and measurement-powered-engine experiments
implement measurement-based feedback and work extraction, but do not use
the first-detection time as the feedback trigger
\cite{Masuyama2018,Naghiloo2018,Dassonneville2026}. Combining these two
ingredients leads naturally to the concept of a quantum first-detection
information engine, in which the first occurrence of a prescribed
measurement outcome triggers a thermodynamic operation in which work is extracted. This engine is the quantum counterpart of the setup considered, e.g., in \cite{Archambault25}.

While a second-law-like inequality for systems subject to a single measurement and subsequent feedback control has been derived in \cite{Sagawa2008}, in \cite{Manikandan2019} an integral fluctuation relation has been derived for a  continuously monitored quantum system in the absence of thermal baths,  where irreversibility is generated by measurement backaction and information acquisition. Such a relation involves an arrow-of-time variable defined through the relative probabilities of forward and time-reversed measurement trajectories, and provides a correction to the integral Jarzynski-like fluctuation theorem, its explicit derivation relying on invertible Kraus operators and on the two-dimensional structure of a qubit.
Here we derive a fluctuation relation for the work extracted by an arbitrary quantum device based on a first-detection mechanism. The correction to the standard Jarzynski equality (JE) is provided by  an experimentally accessible quantity, namely the average first-detection time for the time-reversed first-detection process.
This feature makes our result particularly well suited to applications in quantum information-to-work conversion. Indeed it does not require reconstructing the full statistics of all possible forward and time-reversed measurement trajectories, which is a daunting task even for systems with more than a few degrees of freedom. Instead, it requires only monitoring the response of the \textit{demon}, or equivalently, the feedback controller.

Specifically we derive two  quantum work fluctuation relations for a
first-detection protocol. One relation involves the total work exchanged
during the sequence of stroboscopic measurements and the final feedback
operation, whereas the other concerns the work extracted by the final
operation alone. Jensen's inequality then yields second-law-like bounds on both the total work
cost of the combined measurement-and-feedback protocol and on the extractable work alone. As an additional figure of merit we also introduce the ergotropy averaged over first-detection events, and show that, for the example system considered in this manuscript, it turns out to be a larger upper bound than the one predicted by the quantum fluctuation relations. \alb{We also derive a third relation characterizing the work performed exclusively during the sequence of repeated measurements.}
Our results thus establish  a quantum framework for information engines fuelled by first-detection-time statistics and provide a   comprehensive quantum thermodynamics description  of their operation.

\begin{figure}[h]
	\center
	\includegraphics[width=8cm]{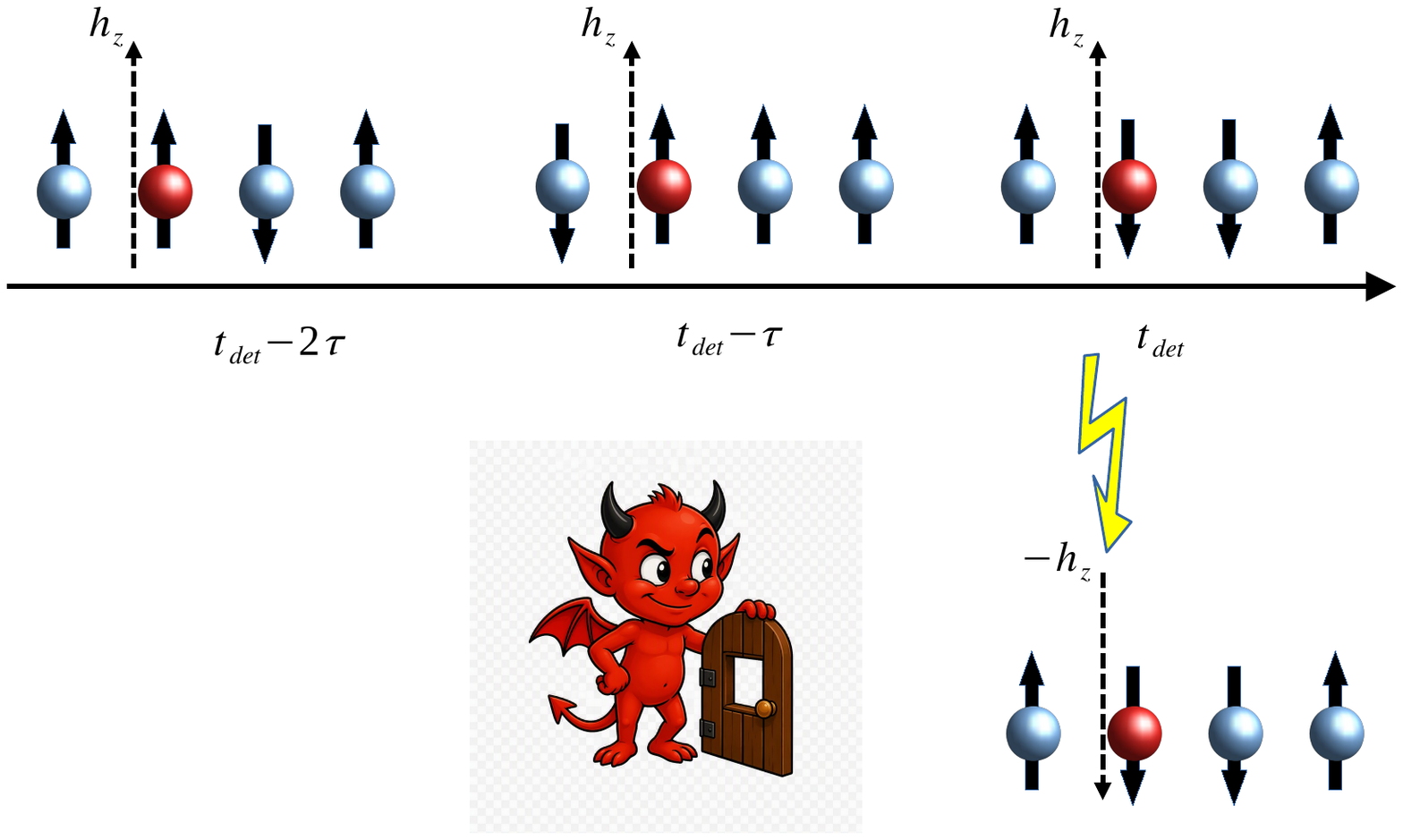}
	\caption{First detection information-to-work converter. The demon measures the $z$-component of the red spin alone with period $\tau$. At time $t_{det}$ the spin is found  for the first time to be antialigned with the external field, and the field acting on the red spin is instantaneously changed by the demon to the value $-h_z$. The operation results in a net extracted work $2 h_z$}
	\label{demon:fig}
\end{figure}

In particular we consider feedback-controlled thermal machines, where the feedback controller, or
 ``demon'', performs repeated  projective measurements on the engine. The demon thus selects  states of high energy, and at their first occurrence quenches the Hamiltonian so that the system is instantaneously brought to a low-energy state, and a positive amount of work is extracted.
 A typical setup is depicted in fig.~\ref{demon:fig}. 
 This protocol resembles the operation of several classical and quantum information engines, in which information acquired through measurement is used by a feedback controller to rectify thermal or quantum fluctuations and convert the resulting non-equilibrium state into work \cite{Toyabe2010, Parrondo2015,Elouard2017,Elouard2018,Masuyama2018,Naghiloo2018,Dassonneville2026,Oftelie2026}. Here, however, the triggering event is the first positive measurement in a stroboscopic setup,  whose classical counterpart has been recently studied for stochastic systems in \cite{Archambault25}.
 
\section{ First-detection processes  }
Here we introduce the concepts and quantities relevant to the present work. For a comprehensive discussion of the quantum first-detection formalism, we refer the reader to Ref.~\cite{Friedman2017}.

Let $\{\ket{x}\}_{x\in\mathcal X}$ be a complete orthonormal set of eigenstates of an observable $X$. Let $\Pi$ be the orthogonal projector onto the subspace spanned by the eigenstates whose eigenvalues belong to a subset $\mathcal X_{\rm det}\subseteq\mathcal X$: 
\begin{equation} \Pi = \sum_{x\in\mathcal X_{\rm det}} \ketbra{x}{x}. \label{Pi:eq}
 \end{equation}

Upon obtaining the positive measurement outcome, the system collapses onto the subspace onto which $\Pi$ projects.

A null measurement projects the system onto the orthogonal complement of the detection subspace, described by
\begin{equation}
\tPi = \II - \Pi.
\end{equation}

Let $U_\tau=\exp(-\ii \tau H_0)$ be  the unitary time--evolution operator, 
with $\hbar=1$ here and in the following, and let $\rho_{in}$ be the initial system state.
With probability $\tp_1=\tr[\tPi  \Ut \rho_{in}\Utd \tPi]$ the first measurement returns a negative outcome, and the post-measurement state of the system will be $r_1=\tPi \Ut \rho_{in} \Utd \tPi/\tp_1$.
After another time interval $\tau$, another negative outcome occurs with probability $\tp_2=\tr[\tPi U_\tau r_1 U^\dagger_\tau \tPi]$, and the state reads $r_2=\tPi U_\tau r_1 U^\dagger_\tau \tPi/\tp_2$.

Thus after $n$ null measurements, the state reads
\begin{equation}
	r_n=(\tPi U_\tau )^n \rho_{in}  (U^\dagger_\tau \tPi)^n \big/\prod_{k=1}^n \tp_k,
	\label{rn:eq}
\end{equation}
with $\tp_n=\tr[\tPi  \Ut r_{n-1}\Utd \tPi]$ and $r_0=\rho_{in}$.
Let us define the operator 
\begin{equation}
	G_\tau = U_\tau \tPi, \label{G:def}
\end{equation} 
we see that the first-detection probability $P_n$, defined as the probability  of a positive measurement outcome on the $n-$th step, after $n-1$ negative outcomes reads
\begin{eqnarray}
	P_n[\rho_{in}]&=&\tr[\Pi \Ut r_{n-1} \Utd \Pi ]\prod_{k=1}^{n-1} \tp_k\nonumber\\
	&=&\tr\pq{\Pi G_\tau^{n-1} U_\tau \rho_{in} U_\tau^\dagger  G_\tau^{\dagger n-1} \Pi}.
	\label{Pn:eq}
\end{eqnarray} 

The  mean first-detection step can thus be calculated as 
\begin{equation}
	\mean{n}_{\rho_{in}}=\sum_{n=1} n P_n[\rho_{in}]\label{eq:n}
\end{equation} 
and the mean first-detection time reads $\tau \mean{n}_{\rho_{in}}$.

A pure state \(|\psi\rangle\) is said to be \emph{dark} with respect to the repeated measurement
protocol if it is
an eigenstate of \(G_\tau\), $G_\tau|\psi\rangle=\lambda |\psi\rangle$,  with an eigenvalue of unit modulus $|\lambda|=1$.
Since \(G_\tau\) is a contraction, this condition implies $\Pi|\psi\rangle=0$,
so that the state never reaches the detection subspace. Equivalently, a
dark state never produces a detection event. 
Besides the trivial case $\tau\to 0$ for which the desired state is never detected because of the Zeno effect, finite values of $\tau$ can result in undetected states for which the average detection time diverges \cite{Friedman2017}.
For an exhaustive discussion of the properties of dark states in, e.g., quantum random walks see \cite{Thiel2020, Wang2024}.

For a general initial state \(\rho_{in}\), we define the survival probability
after \(n\) measurement attempts as
\[
S_n[\rho_{in}]
=
\tr\left[
G_\tau^n U_\tau \rho_{\rm in} U_\tau^\dagger (G_\tau^\dagger)^n
\right]=1-\sum_{k=1}^n P_k[\rho_{in}]
\]
The sequence \(\{S_n\}_{n\ge0}\) is monotonically decreasing and its
asymptotic value $S_\infty :=\lim_{n\to\infty} S_n$
measures the fraction of the initial state that survives indefinitely.


We say that a general initial state $\rho_{in}$ contains a dark component whenever
$S_\infty[\rho_{in}]>0$.
In particular, $\rho_{in}$ is completely dark if $S_n=1$ for all $n$. In general a mixed state may contain both detectable and dark components.

\alb{In the following we assume that the initial states satisfy $S_{\infty}[\rho_{in}]=0$: this condition ensures that the engine is eventually activated with
probability one, so as work can be extracted.}


\section{ Quantum JE}
We now review the quantum Jarzynski equality (QJE) \cite{Tasaki2000, Kurchan2000,Talkner2007, Funo2018} in order to establish the necessary background for the rest of the paper.

We consider a system driven by a time-dependent Hamiltonian $\mathcal{H}(t)$, generating the unitary evolution $ \cU_{t}$ from the initial time $t_0=0$ to the final time $t_1$. We define $\mathcal{H}_0 \equiv \mathcal{H}(0)$ and $\mathcal{H}_1 \equiv \mathcal{H}(t_1)$ as the initial and final Hamiltonians, respectively.
If  the system is initially in the eigenstate $\mathcal{E}_{0,m_0}$ of the initial Hamiltonian $\mathcal{H}_0$, and denoting by $\mathcal{E}_{1,m_1}$ the eigenstates of the final Hamiltonian, the probability distribution of the work done on the system reads
\begin{equation}
	P(W|m_0)=\sum_{m_1} \delta(\mathcal{E}_{1,m_1}-\mathcal{E}_{0,m_0}-W) |\bra{\mathcal{E}_{1,m_1}} \mathcal{U}_{t_1}\ket{\mathcal{E}_{0,m_0}}|^2.
\label{PW0:eq}
\end{equation}
Assuming that the system is initially in the thermal equilibrium state $\rho_0=\exp(-\beta \mathcal{H}_0)/Z_0$, one finds the quantum JE
\begin{equation}
	\mean{\E^{-\beta W}}=\int \D W \E^{-\beta W}\sum_{m_0} P(W|m_0) \frac{\E^{-\beta \mathcal{E}_{0,m_0}}}{Z_0}=\E^{-\beta \Delta F},
\label{JE:eq0}
\end{equation}
where $\Delta F=-k_B T \log\p{Z_1/Z_0}$, and $Z_i=\tr[\exp(-\beta \cH_i)]$.


The above derivation of the quantum Jarzynski equality for an isolated driven quantum system is based on the two-projective-measurement (TPM) scheme \cite{ Kurchan2000,Talkner2007}, and is consistent with the quantum first-detection formalism of Refs.~\cite{Friedman2017,Thiel2018}, in which the system is subjected to repeated stroboscopic projective measurements.

\section{ First-detection engines }
As discussed in the introduction, in this manuscript we consider information-fuelled machines operating through a single-stroke protocol. When the measurement outcome associated with the projector $\Pi$ is obtained, the system Hamiltonian is instantaneously changed from $H_0$ to $H_1$, with  the two Hamiltonians and the projector chosen such that $\tr[\Pi (H_0-H_1)\Pi]\ge0$.


For control driven by the first-detection time, one has to evaluate the  extracted work averaged over the detection events.
The state of the system after the $n$-th positive outcome, following $n-1$ negative outcomes is $r_n$ (eq.~\eqref{rn:eq}), and the first-detection probability $P_n$ is given by eq.~\eqref{Pn:eq}.

For an observable $A$ we now introduce the first-detection expectation value (or detection--conditioned expectation value)
\begin{eqnarray}
	\mean{A}_{\det}&=&\sum_n \tr[A\,  \Pi \Ut r_{n-1} \Utd \Pi ]\prod_{k=1}^{n-1} \tp_k\nonumber\\
	&=&\sum_n \tr[A\,  \Pi G_\tau^{n-1} U_\tau \rho_{in} U_\tau^\dagger  G_\tau^{\dagger n-1} \Pi]\nonumber\\&=&\sum_n P_n[\rho_{in}] \tr[A \, \rho_{n|\det}], 	\label{A:rhodet}
\end{eqnarray}
i.e., the expectation value of $A$ over the ensemble of first-detection events.

We can also rewrite this quantity as 
\begin{equation}
	\mean{A}_{\det}=\sum_n P_n \tr[A \, \rho_{n|\det}],
\end{equation} 
where 
\begin{equation}
\rho_{n|\det}=
\Pi G_\tau^{n-1} U_\tau \rho_{in} U_\tau^\dagger
(G_\tau^\dagger)^{n-1}\Pi/P_n[\rho_{in}] ,\label{rho:post:norm}
\end{equation}
 is the normalized post-measurement state associated with first-detection at the $n$-th measurement.

In the following we make the same assumption as in the JE, namely that the system is initially in a thermal equilibrium state, characterized by the inverse temperature $\beta$ and the initial Hamiltonian $H_0$, $\rho_{in}=\rho_0=\exp(-\beta H_0)/Z_0$. 
Analogously to the QJE, if the system is initially in the eigenstate $\ket{\epsilon_{0,k_0}}$ of $H_0$, and is found in the  eigenstate $\ket{\epsilon_{1,k_1}}$ of $H_1$ after a positive measurement determined by $\Pi$, the total energy change for this specific outcome is $W_{\mathrm{tot}}=\epsilon_{1,k_1}-\epsilon_{0,k_0}$. The other quantity of interest is the energy change following a positive measurement and the subsequent quench $H_0\to H_1$.
Thus, we first consider  two fluctuating quantities that characterize the thermodynamic performance in terms of the total energy change and extracted work: their values averaged over the detection events read, respectively, 
\begin{eqnarray}
\wtot&=&\mean{H_1}_{\det}-\mean{H_0}_0 =\mean{H_1}_{\det}-\tr[H_0 \rho_0],\label{wtot:eq}\\
-\wex&=&\mean{H_1-H_0}_{\det}. \label{wex:eq} 
\end{eqnarray} 
\alb{where we have used the lowercase notation for $\wtot$ and $\wex$ to indicate the expectation values as opposed to the fluctuation quantities which we indicate with the uppercase notation.}
The first fluctuating quantity is analogous to the total energy change that enters in the Jarzynski equality \eqref{JE:eq0}. However, in the setup  considered here  the total energy change of the machine also includes the thermodynamic cost of the measurement strokes.
The second quantity  measures the  work extracted by the demon upon detection of the triggering states, and in order for the machine to extract work one requires the average over the detection events to be positive $\wex>0$. 
Notice that having $\wex>0$ does not entail the inequality $\wtot<0$, as measuring can indeed be expensive in terms of work performed on the machine.

We will derive fluctuation relations and bounds for both quantities.

Let us first introduce the time reversal operator $\theta$ satisfying the antilinearity relation 
$-\ii (\theta H \theta^{-1}) \theta \ket \psi=\theta \ii H \ket  \psi$.
If $H$ is invariant under time reversal, $\theta H \theta^{-1}=H$, then one finds $-\ii H \theta \ket \psi=\theta \ii H \ket  \psi$, or $[H,\theta]=0$ \cite{sakurai_napolitano_2017}.
Let $\tilde A=\theta A \theta^{-1}$ denote the time-reversed operator (or state).
Let us define the normalized state $\varrho_1$ and its time reversed
\begin{equation}
 	\varrho_1=\frac{ \Pi \E^{-\beta  H_1} \Pi}{\cZ_1},\qquad	\trhou=\frac{\tilde \Pi \E^{-\beta \tilde H_1} \tilde \Pi}{\cZ_1}, \label{varrho1:eq}
\end{equation}
with $ \cZ_1=\tr[ \Pi \E^{-\beta  H_1}  \Pi]=\tilde \cZ_1$. We also introduce the free energy difference $\Delta \cF=-k_B T \ln(\cZ_1/Z_0)$.

We now state our first result which is the first-detection counterpart of the quantum JE \eqref{JE:eq0}, namely a fluctuation relation for the total  energy  change, whose expected value is given by eq.~\eqref{wtot:eq}. The following equality holds
\begin{equation}
	\mean{\E^{-\beta W_{\mathrm{tot}}}}_{\det}=\frac{\cZ_1}{Z_0} \sum_{n=0}^{\infty}S_n[\trhou]=\E^{-\beta (\Delta \cF -T \log\mean{\tilde n}_{\trhou})} \label{JE:wtot}
\end{equation}
where 
 $\mean{ \tilde n}_{\trhou}$ is the mean first-detection step of the process with $\trhou$ as non-equilibrium initial state along a time-reversed dynamics generated by the unitary 
 $\mathrm{exp}(-\ii \tau \tilde H_0)$ and the projection operator $\tilde \Pi$. Here and in the following we set $k_B=1$.
The above work fluctuation relation establishes a  nontrivial  relation between the work statistics along the forward dynamics and the survival statistics along the backward dynamics.
The proof of eq.~\eqref{JE:wtot} is given in the appendix. \alb{We notice that being a projection onto the detection subspace, $\trhou$ \eqref{varrho1:eq}  contains no dark component so as $S_\infty[\trhou]=0$.}  

Setting $\Pi=\II$,  i.e. making the detector click with certainty after the first time interval $\tau$ independently of the system state, one immediately finds   $\mean{\tilde n}_{\trhou}=1$ and $\cZ_1=Z_1=\tr[ \exp(-\beta H_1) ]$ and recovers the standard JE \eqref{JE:eq0}.

Thus the information that one gathers about the system through stroboscopic measurements enters the fluctuation relation as a thermodynamic contribution lowering the free energy difference that appears in the standard QJE.
Indeed, one can interpret the term $-T \ln \mean{\tilde n}_{\trhou}\le 0$ as an entropic term adding up to the free energy difference $\Delta \cF$ and arising from the detection protocol. 

Application of Jensen's inequality to eq.~\eqref{JE:wtot} leads to a lower bound for $\wtot$, 
\begin{equation}
\wtot=\mean{H_1}_{\det}-\mean{H_0}_{0}\ge \Delta \cF -T  \log\mean{\tilde n}_{\trhou}.
\label{ineq1}
\end{equation} 
We now proceed to derive the fluctuation relation for the extracted work, whose expected value is given by eq.~\eqref{wex:eq}.
Assuming that the spectrum of the operator $\Delta H=H_1-H_0$ is bounded below, which is the case for, e.g., a spin system,
and introducing the shifted extracted work, with average $w'_{ex}=-\wex-\mean{H_0}_0=\mean{H_1-H_0}_{\det}-\mean{H_0}_0$,
 the following fluctuation relation holds for the corresponding fluctuating quantity 
\begin{equation}
	\mean{\E^{-\beta W'_{ex}}}=\frac{\cZ_{\Delta}}{Z_0} \sum_{n=0}^{\infty}S_n[\tilde\varrho_{\Delta}]
		=\E^{-\beta (\Delta \cF_{\Delta} -T \log\mean{\tilde n}_{\tilde\varrho_{\Delta}})} \label{JE:wex}
\end{equation}
where
\begin{equation}
	\trhoD=\frac{\tilde \Pi \E^{-\beta (\Delta \tilde H)} \tilde \Pi}{\cZ_{\Delta}}, \label{varrho10:eq}
\end{equation}
is the out-of-equilibrium initial state for the time-reversed dynamics generated by the unitary $\mathrm{exp}(-\ii \tau \tilde H_0)$ and the projection operator $\tilde \Pi$, 
with $ \cZ_{\Delta }=\tr[ \Pi \E^{-\beta \Delta H}  \Pi]=\tilde \cZ_{\Delta}$ the corresponding partition function. The free energy difference reads $\Delta \cF_{\Delta}=-k_B T \ln(\cZ_{\Delta}/Z_0)$, and  $\mean{\tilde n}_{\tilde\varrho_{\Delta}}$ 
is the mean first-detection step along the time-reversed process with $\tilde\varrho_{\Delta}$ as initial state.

Applying Jensen's inequality to eq.~\eqref{JE:wex}
one obtains the upper bound
\begin{eqnarray}
w_{ex}&=&-(\mean{H_1-H_0}_{\det})\nonumber \\
&\le&-(\Delta \cF_{\Delta } -T \log\mean{\tilde n}_{\tilde\varrho_{\Delta}}) -\mean{H_0}_{0} \label{wex:Jens}
\end{eqnarray} 
where the log of the mean detection step appears again as an entropic correction term to the free energy difference.

\alb{Given that the quantity $\wtot$ accounts for the combined change in energy due to the measurements and the mechanical operation, it is of interest to consider the fluctuation relation for the energy involved in the measurement process alone, whose value averaged over the detection events reads 
\begin{equation}
	\wmeas=\mean{H_0}_{\det}-\mean{H_0}_0,
\end{equation}
this quantity corresponding to the average energy change in the system undergoing the stroboscopic measurements immediately {\it before} the mechanical operation $H_0\to H_1$ is implemented by the demon at the detection event.
The fluctuation relation for the corresponding fluctuation quantity reads 
\begin{equation}
	\mean{\E^{-\beta W_{\mathrm{m}}}}_{\det}=\E^{-\beta (\Delta \cF_\mathrm{m} -T \log\mean{\tilde n}_{\trhoz})}, \label{JE:meas}
\end{equation}
where the out-of-equilibrium initial state for the  time-reversed process reads $\trhoz=\tilde \Pi \mathrm{exp}(-\beta  \tilde H_0) \tilde \Pi/\cZ_{\mathrm{ 0}}$, with $\cZ_{\mathrm{ 0}}= \mathrm{tr}[\tilde \Pi \mathrm{exp}(-\beta  \tilde H_0) \tilde \Pi]$ and $\Delta \cF_\mathrm{m}=-k_B T \ln(\cZ_0/Z_0)$.
Application of Jensen's inequality to eq.~\eqref{JE:meas} leads to the lower bound for the energy change due to the measurement process alone
\begin{equation}
\wmeas=\mean{H_0}_{\det}-\mean{H_0}_{0}\ge \Delta \cF_{\mathrm m} -T  \log\mean{\tilde n}_{\trhoz}.
\label{ineq2}
\end{equation} 
}
In order to illustrate the theoretical results of this paper and to gain physical insight
into their implications we apply them to spin-1/2 systems characterized by Pauli operators
$\sigma^{x,y,z}$.

We consider a two-spin system,  with Hamiltonian $H_0=-J \sigma_{a}^x  \sigma_{b}^x - h (\sigma_{a}^z + \sigma_{b}^z)$, with $J,\, h>0$. 
As a target state we consider the state in which the spin $b$ is  down, characterized by the projector $\Pi=\II_2 \otimes \ket{\downarrow}\bra{\downarrow}$, and where $\II_2$ is the $2\times2$ identity matrix. Immediately after a positive measurement outcome, the demon changes the sign of the field on the $b$ spin, with the Hamiltonian now reading $H_1=-J \sigma_{a}^x  \sigma_{b}^x - h (\sigma_{a}^z - \sigma_{b}^z)$ and a net extracted work $w_{ex}=2 h >0$.
We do not consider the case $\Pi= \ket{\downarrow \downarrow}\bra{\downarrow \downarrow}$ as for this choice the operator $G_\tau$ has two dark states $(\ket{\uparrow \downarrow} \pm \ket{\downarrow \uparrow})/\sqrt 2 $.


\begin{figure}[h]
	\center
	\includegraphics[width=8cm]{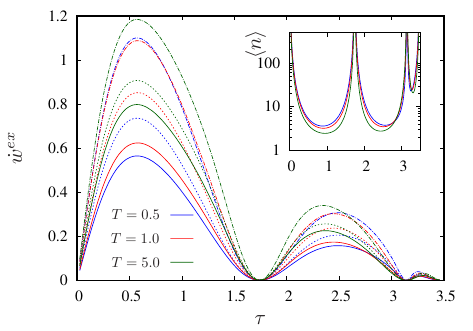}
	\caption{Full lines: average extracted work rate $\dot w_{ex}$ (eq.~\eqref{eq:dwex}) as a function of the sampling interval $\tau$ for the two--spin model described in the text, and different values of the temperature. Dashed lines:  upper bound for the extracted work rate $\dot w^{ex}$ as determined by the upper bound for $ w_{ex}$ given in eq.~\eqref{wex:Jens}. Dash-dotted lines: first-detection mediated ergotropy per time unit $\mean{\mathcal{E}}_{\mathrm{det}}/(\tau \mean{n}_{\rho_0})$ as a function of $\tau$, with $\mean{\mathcal{E}}_{\mathrm{det}}$ defined in the text. 
		Inset: Average first-detection time $\mean{n}_{\rho_0}$, as defined in equation \eqref{eq:n}, as a function of $\tau$. Diverging values of $\mean{ n}_{\rho_0}$ indicates that for some value of $\tau$ the desired state remains undetected, a feature found in other systems subject to stroboscopic measurements \cite{Varbanov2008, Friedman2017}.}
	\label{dWave:fig}
\end{figure}

For stochastic and quantum devices operating with single or multiple strokes, one is interested in the output power rather than the output work. We therefore define an average output power as 
\begin{equation}
	\dot w_{ex}= \frac{\wex}{\tau \mean{n}_{\rho_0}},\label{eq:dwex}
\end{equation}
whose numerator is bounded by eq.~\eqref{wex:Jens}.
\alb{
We see that, in order for the above quantity to be well defined and nonzero, one must impose the additional condition $\mean{n}_{\rho_0}<\infty$ on the forward process. The requirement  $S_{\infty}[\rho_{in}]=0$, which ensures that the engine is activated with probability one and work is thus extracted, is necessary for the mean first-detection step to be finite and, for a finite-dimensional Hilbert space, it is also a  sufficient condition \cite{Thiel2020}.}

As a further figure of merit for  evaluating the performance of a device based on stroboscopic measurements and first-detection events that trigger the feedback circuit, we consider the ergotropy,  defined as the maximum extractable energy from a system with Hamiltonian $H$ and in a state $\rho$ with a cyclic
unitary process \cite{Allahverdyan2004}
\begin{equation}
	\mathcal E =	\mathcal E(\rho, H)=\tr[H \rho]-\min_U \tr[H U\rho U^\dagger].
\end{equation}
 In fact, for the simple example considered here, namely  flipping the external magnetic field along $z$ is equivalent to a rotation around the $x$-axis generated by the unitary $\sigma_\alpha^x$, but more complex rotations (or translations for systems with spatial coordinates) are described by unitary operators too. 
The state conditioned on first-detection at step $n$ is given by eq.~\eqref{rho:post:norm}.
Thus one can introduce an ergotropy averaged over the first-detection events
\begin{equation}
	\mean{\mathcal E}_{\det}=\sum_n  P_n\,  \mathcal E(\rho_{n|\det},H_0). \label{eq:ergo:av}
\end{equation}
The results for the extracted power \eqref{eq:dwex} as a function of $\tau$, its upper bound as determined by the Jensen's inequality \eqref{wex:Jens} and the averaged ergotropy per time unit are shown in fig.~\ref{dWave:fig} for different values of the temperature. The bound for the power  obtained through the ergotropy turns out to be an upper bound for the other two as expected.
Inspection of fig.~\ref{dWave:fig} indicates that the maxima of $\dot w_{ex}$ closely follow  the minima of $\mean{n}_{\rho_0}$, the former occurring at a slightly smaller value of $\tau$.

Thus far, we have implicitly assumed that the device is a closed quantum system. However, the result of the present paper can be generalised to the case where the system of interest ($S$) is connected to a bath ($B$), with total Hamiltonian $H_0=H_0^{(s)}+H^{(B)}+V$ where $V$ is an interaction term between $S$ and $B$. If  the quench is performed on the system of interest alone $H_0\to H_1=H_1^{(s)}+H^{(B)}+V$, and the projector ~\eqref{Pi:eq} acts only on $S$, i.e., the demon monitors only an observable of the device, the quantum fluctuation relations \eqref{JE:wtot} and \eqref{JE:meas} hold without modification.  Extending eq.~\eqref{JE:wex} requires more caution, and one has to require at least that the Hilbert space of the bath is finite-dimensional given that, in this case, one deals with the exponential of a system operator alone $H_1-H_0=H_1^{(s)}-H_0^{(s)}$. This issue, which requires a rather technical analysis,  will be fully addressed in a future work. 

\section{ Conclusions}
In this paper, we have studied the work statistics of quantum information engines based on a first-detection protocol. The resulting fluctuation relations, and hence the resulting bounds on both the total work associated with the measurement and mechanical operations and the extracted work alone, depend on an experimentally accessible quantity, namely, the mean first-detection step. The correction to the standard Jarzynski equality takes the form of an entropic contribution that reduces the effective free-energy cost and quantifies the information acquired about the system through repeated measurements. By introducing a first-detection ensemble, we also define and evaluate a first-detection-averaged ergotropy, which provides an independent upper bound on the extracted work. Using a practical example, we show that the bound derived from the fluctuation relation is tighter than the ergotropy bound. This upper bound on the extracted power exhibits a nontrivial dependence on the sampling interval and therefore provides a control parameter for optimizing the power output.
\alb{Recent experiments provide concrete examples of information-to-work conversion that could be adapted to the scheme presented in this paper, and thus convert the energy released by a first-detection-triggered quench into useful work.
For example  in Ref.~\cite{Dassonneville2026}, the engine is a superconducting
qubit powered by repeated quantum measurements and measurement-induced
energy is transferred to a coherent microwave field and used to amplify an
incoming signal. In Ref.~\cite{Zhang2025}, the information engine consists of the internal
two-level degree of freedom of a single trapped ion, while the ion's
quantized motional mode receives and stores the extracted energy as a
quantum battery.}

\acknowledgments
The author is grateful to Tommaso Roscilde for an inspiring discussion.

\renewcommand{\theequation}{A\arabic{equation}}
\renewcommand{\thefigure}{A\arabic{figure}}

\renewcommand{\thesection}{A.\arabic{section}}

\setcounter{equation}{0}
\setcounter{section}{0}

\onecolumngrid
~
\clearpage

\begin{center}
	\large{\bf Appendix}
\end{center}

~
\twocolumngrid

\section{Work distribution and proof of the quantum work fluctuation relations}
In order to establish eq.~\eqref{JE:wex} we need to introduce the work PDF in the case of repeated measurements, analogous to the work PDF in absence of intermediate measurements \eqref{PW0:eq}.
Let us assume that the system at $t=0$ is in the eigenstate  $\ket{\epsilon_{0,k_0}}$ of the Hamiltonian $H_0$.
The probability of being in the eigenstate $\epsilon_{1,k_1}$ of the Hamiltonian $H_1$, conditioned on detection at step $n$ reads 
\begin{equation}
	|\bra{\epsilon_{1,k_1}} \Pi G_\tau^{n-1} \ket{\epsilon_{0,k_0}}|^2/P_n[\ket{\epsilon_{0,k_0}}].
\end{equation}
The total energy change for this specific outcome reads $W_{\mathrm{tot}}=\epsilon_{1,k_1}-\epsilon_{0,k_0}$.

Thus, assuming that the system is initially at equilibrium at temperature $\beta$ with Hamiltonian $H_0$, $\rho_0=\mathrm{exp}(-\beta H_0)/Z_0$, the PDF of the work averaged over the detection events reads
\begin{eqnarray}
	P_{\det}(W_{\mathrm{tot}})&=&\sum_{k_0,k_1} \delta(W_{\mathrm{tot}} - (\epsilon_{1,k_1} -\epsilon_{0,k_0}) )\nonumber \\ 
	&\times& \sum_{n=1}^\infty|\bra{\epsilon_{1,k_1}} \Pi G_\tau^{n-1} \ket{\epsilon_{0,k_0}}|^2 \frac{\e^{-\beta  \epsilon_{0,k_0}}}{Z_0}\nonumber \\ 
	&&\label{pwtot:eq}.
\end{eqnarray}
Next, we evaluate the following expectation value
\begin{eqnarray}
	&&\mean{\e^{-\beta W_{\mathrm{tot}}}}_{\det}=\int \D  W_{\mathrm{tot}} \e^{-\beta W_{\mathrm{tot}}} \, P_{\det}(W_{\mathrm{tot}})\label{meaneW}\\
	&&= \sum_{k_1,k_0} \frac{\e^{-\beta  \epsilon_{1,k_1}}}{Z_0} \sum_{n=1}^\infty|\bra{\epsilon_{1,k_1}} \Pi G_\tau^{n-1} \ket{\epsilon_{0,k_0}}|^2 \label{meaneW1}\\
	&&=  \sum_{k_1} \frac{\e^{-\beta  \epsilon_{1,k_1}}}{Z_0} \sum_{n=1}^\infty \tr[G_\tau^{\dagger n-1}\Pi \ket{\epsilon_{1,k_1}}\bra{\epsilon_{1,k_1}}\Pi G_\tau^{n-1}].\nonumber \\
	&& \label{meaneW2}
\end{eqnarray}

We now define
\begin{equation}
f_n=\Tr[\E^{-\beta H_1} \Pi G_\tau^{n-1} \Ut \E^{\beta H_0} \rho_0 \Utd  G_\tau^{\dagger n-1} \Pi]\label{eq:fn0}, 
\end{equation}
and notice that eqs.~\eqref{meaneW}--\eqref{meaneW2} can be written as
\begin{eqnarray}
	\mean{\e^{-\beta W_{\mathrm{tot}}}}_{\det}= \sum_{n=1}^\infty f_n, \label{mean:corr} 
\end{eqnarray}

We remind the reader of the definition of $G_\tau$ eq.~\eqref{G:def}, 
and of  $\varrho_1$ and $\trhou$ given in  eq.~\eqref{varrho1:eq},
and notice that the following equalities hold 
\begin{eqnarray}
	&&f_n=\frac {\cZ_1}{ Z_0}  \tr[G_\tau^{\dagger n-1} \Pi \varrho_1 \Pi
	G_\tau^{n-1}]\nonumber\\
	&&=f_{n-1} -\frac {\cZ_1}{ Z_0}\tr[\Pi \p{\Utd \tPi}^{n-2} \Utd \varrho_1\Ut
	\p{\tPi \Ut}^{n-2}\Pi ].\nonumber\ \\
	&& \label{fn:recurs}
\end{eqnarray} 
We now focus on the second term of the last equality, and by using  the relations $\theta \Ut \theta^{-1}=\tUtd$, $\theta \Utd \theta^{-1}=\tUt$ we notice that 
\begin{eqnarray}
	&&\tr[\Pi \p{\Utd \tPi}^{n-2} \Utd \varrho_1 \Ut
	\p{\tPi \Ut}^{n-2}\Pi ]\nonumber\\
	&&= \tr[\tilde \Pi \p{\tUt \ttPi}^{n-2} \tUt \trhou \tUtd \p{\ttPi \tUtd}^{n-2}\tilde \Pi ]=\tilde P_{n-1}[\trhou]\nonumber\label{Ptrho1:eq}\\
	&&
\end{eqnarray} 
where we have used eq.~\eqref{Pn:eq} to define the probability of first-detection at $n-1$--th step with time-reversed dynamics generated by $\tilde U_\tau$, time-reversed projector $\tilde \Pi$, and initial non--equilibrium probability distribution $\trhou$.
We notice that, if the Hamiltonians $H_0$ and $H_1$  and the projector onto the detection  subspace $\Pi$ are invariant under time reversal, the rhs of eq.~\eqref{Ptrho1:eq} becomes $P_{n-1}[\varrho_1]$.
Given that $f_1=\cZ_1/Z_0$, eq.~\eqref{fn:recurs} becomes
\begin{eqnarray}
	f_n=\frac{\cZ_1}{Z_0}(1-\sum_{k=1}^{n-1} \tilde P_{k}[\trhou])=\frac{\cZ_1}{Z_0} S_n[\trhou]\label{fn:simp},
\end{eqnarray} 
thereby establishing a relation between the work statistics along the forward dynamics, and the survival probability along the backward dynamics.

Considering again eq.~\eqref{mean:corr}, and using eq.~\eqref{fn:simp}, we notice that 
\begin{eqnarray}
	&&\sum_{n=1}^\infty f_n=\frac{\cZ_1}{Z_0} \sum_{n=1}^\infty (1-\sum_{k=1}^{n-1}\tilde P_{k}[\trhou])=\frac{\cZ_1}{Z_0} \sum_{n=1}^\infty \sum_{k=n}^\infty  \tilde P_{k}[\trhou]\nonumber\\
	&&=\frac{\cZ_1}{Z_0}\sum_{k=1}^\infty  k \tilde P_{k}[\trhou]=\frac{\cZ_1}{Z_0} \mean{\tilde n}_{\trhou},\label{eq:fn:final}
\end{eqnarray}
which proves eq.~\eqref{JE:wtot} in the main text.

\alb{We now emphasize the connection between quantum fluctuation relations and two-time correlation functions.}
We start by  noticing that the QJE \eqref{JE:eq0}  can be expressed in the more compact form as 
\cite{DeChiara2022}
\begin{equation}
\mean{\E^{-\beta W}}=\mean{\E^{-\beta \cH(t_1)} \E^{\beta \cH(0)}}=\E^{-\beta \Delta F},
\label{JE:eq}
\end{equation} 
 where we have introduced the two-time correlation for two observables $A$ and $B$, which in the Heisenberg picture reads
\begin{equation}
\mean{A(t)B(0)}=\tr[\cU_t^\dagger A\cU_t B\rho_0],
\label{AB:eq}
\end{equation} 
with $A=\exp(-\beta \cH_1)$ and  $B=\exp(+\beta \cH_0)$ for the specific case at hand.

Next we introduce the analogue of a two-time correlation function \eqref{AB:eq} for the
first-detection ensemble, and replace the ordinary unitary
evolution in eq.~\eqref{AB:eq} by the survival--conditioned evolution
\begin{equation}
	\mean{A(t_{\det})B(0)}_{\det}=\sum_n \tr[A\,  \Pi G_\tau^{n-1} U_\tau B \rho_{in} U_\tau^\dagger  G_\tau^{\dagger n-1} \Pi].\label{ABdet:eq}
\end{equation}

Using  the definition of $f_n$, eq.~\eqref{eq:fn0}, and  the two-time correlation for the first-detection ensemble, eq.~\eqref{ABdet:eq}, we see that eq.~\eqref{mean:corr}, and thus the quantum fluctuation relation introduced in this manuscript, can be recast in the suggestive form resembling a two-time correlation function 
\begin{equation}
	\mean{\e^{-\beta W_{\mathrm{tot}}}}_{\det}=\sum_{n=1}^\infty f_n=\mean{\E^{-\beta H_1 (t_{\det})} \E^{\beta H_0(0)}}_{\det}.
\end{equation}

The proof of eq.~\eqref{JE:wex} follows by replacing the Hamiltonian $H_1$ with $\Delta H$ in eq.~\eqref{eq:fn0}
and subsequent equations. In particular in eq.~\eqref{eq:fn:final} $\trhou$ has to be replaced by $\tilde\varrho_{\Delta}$ and  $\cZ_1$ has to be replaced by $\cZ_\Delta$.

Similarly the proof of eq.~\eqref{JE:meas} follows by replacing the Hamiltonian $H_1$ with $H_0$ in eq.~\eqref{eq:fn0}
and subsequent equations. 

\bibliography{bibliography}
\end{document}